\documentclass[aps,twocolumn,prd,showpacs,showkeys,preprintnumbers,superscriptaddress,nobibnotes,floatfix,longbibliography,nofootinbib]{revtex4-1}

\pdfoutput=1

\usepackage[english]{babel}
\usepackage{amsmath,amssymb,amsbsy,amstext,amsthm,booktabs,exscale,relsize,slashed,graphicx,amsfonts,upgreek,xcolor,amsmath}
\usepackage{multirow,dcolumn,bm,enumerate,url,hyperref}

\usepackage{latexsym}
\usepackage{feynmp}
\usepackage{color}
\usepackage{cancel}
\usepackage{multirow,array}
\usepackage{float}
\usepackage{subfigure}
\usepackage{mathtools}
\usepackage{graphicx}
\usepackage{soul}

\newcommand{\gev}{{\rm GeV}}

\newcommand{\cm}{{\rm cm}}
\newcommand{\km}{{\rm km}}
\newcommand{\s}{{\rm s}}

\newcommand{\fiv}{f_{\rm IV}}

\begin{document}

\title{Opening windows with Isospin-Violating Dark Matter}

\author{Jason Kumar}
 \thanks{{\scriptsize Email}: \href{mailto:jkumar@hawaii.edu}{jkumar@hawaii.edu}}
\affiliation{\smaller Department of Physics and Astronomy, University of Hawai'i, Honolulu, HI 96822, USA}

\author{Danny Marfatia}
 \thanks{{\scriptsize Email}: \href{mailto:dmarf8@hawaii.edu}{dmarf8@hawaii.edu}}
\affiliation{\smaller Department of Physics and Astronomy, University of Hawai'i, Honolulu, HI 96822, USA}

\author{Ningqiang Song}
\thanks{{\scriptsize Email}: \href{mailto:songnq@itp.ac.cn}{songnq@itp.ac.cn}}
\affiliation{\smaller Institute of Theoretical Physics, Chinese Academy of Sciences, Beijing, 100190, China}

\begin{abstract}
We consider the effect of isospin-violating dark matter-nucleon interactions on direct 
detection constraints in the regime of small dark matter mass and large scattering cross section.
Isospin-violation can lead to both reductions in sensitivity (due to a reduced cross section for 
scattering with nuclei in the detector) and enhancements in sensitivity (due to a reduced cross section 
for scattering in the overburden).  Isospin-violating effects can thus open up some closed regions of 
parameter space, while closing off other regions.  
\end{abstract}

\maketitle 

\section{Introduction}
\label{sec:intro}

In recent years, it has been appreciated that if
dark matter (DM) has a relatively large cross section for scattering with nuclei, then direct detection experiments may be shielded from the DM flux by the atmospheric and terrestrial overburden~\cite{Davis:2017noy,Kavanagh:2017cru,Hooper:2018bfw,Emken:2018run,Neufeld:2018slx,Emken:2019tni,Bramante:2022pmn}. 
Thus far, studies of this scenario have largely 
focused on the case in which dark matter spin-independent (SI) interactions couple equally 
to protons and neutrons.  In this work, we consider the case of isospin-violating 
dark matter (IVDM)~\cite{Kurylov:2003ra,Giuliani:2005my,Chang:2010yk,Feng:2011vu} with a large SI scattering cross section.

Isospin-violating interactions can lead to a reduced scattering cross section with 
large nuclei, since interactions with protons and neutrons can partially cancel each 
other.  This can lead to two competing effects on the sensitivity of a direct detection 
experiment: the sensitivity can be reduced due to a reduced cross section for dark matter particles $\chi$ to 
scatter with the detector material, but it can also be enhanced due to a reduction in 
dark matter scattering with the overburden.  We will determine how the allowed DM parameter space in the $(m_\chi,\sigma)$-plane (where $\sigma$ is the DM-proton scattering cross section $\sigma_{\chi p}$ or the DM-neutron scattering cross section $\sigma_{\chi n}$) changes for the case of isospin-violating SI scattering, as 
the ratio of DM couplings to neutrons and to protons is varied.  

We focus on relatively low DM mass ($0.1~\gev< m_\chi < 10~\gev$), for which direct detection 
experiments often have suppressed sensitivity, and for which there are greater prospects for isospin-violation 
to alter the allowed regions of parameter space.  For light DM, scattering with nuclei can 
significantly change the direction of dark matter particle motion.  The commonly used straight-line approximation 
(SLA) for the dark matter path through the overburden may not be valid in this case.  We instead use 
an analytical approach, described in Ref.~\cite{Cappiello:2023hza}, which accounts for changes in the path of the DM particle
due to scattering, statistical 
fluctuations in the number of scatters in the overburden, and the energy loss per scatter.

The plan of this paper is as follows.  We describe the effects of isospin violation on dark matter scattering 
in the detector and in the overburden in Section~\ref{sec:IVDMeffects}.  In Section~\ref{sec:IVDMComp} 
we describe the computation of the velocity distribution at the detector.  In Section~\ref{sec:IVDMExp}, we 
obtain constraints on IVDM.  We conclude in Section~\ref{sec:Conclusion}.

\section{Isospin-violating Interactions}
\label{sec:IVDMeffects}

With isospin-violating interactions, dark matter may couple differently to protons and neutrons. Assuming spin-independent interactions, the differential scattering cross section with a nucleus $j$ is~\cite{Feng:2011vu}
\begin{equation}
    \dfrac{d\sigma_{j}}{dE_R}=\dfrac{m_{A_j}\sigma_0}{2\mu_N^2v^2}\fiv^2 F_j^2(E_R)\,,
\end{equation}
where $m_{A_j}$ is mass of the nucleus with mass number $A_j$, $\mu_N$ is the reduced mass of the $\chi$-nucleon system, $\sigma_0$ is the $\chi$-nucleon scattering cross section and $F_j^2(E_R)$ is the nuclear form factor. The isospin-violating effects are encoded in the different couplings to protons $f_p$ and neutrons $f_n$, and we have defined 
\begin{equation}
\fiv\equiv f_pZ_j+f_n(A_j-Z_j)\,,
\end{equation}
where $Z_j$ is the atomic number of nucleus $j$. For comparison with various experimental limits, we normalize the couplings so that either $f_p=1$ and $\sigma_0\equiv \sigma_{\chi p}$, or $f_n=1$ and $\sigma_0\equiv\sigma_{\chi n}$. The expected number of events at direct detection experiments is
\begin{equation}
\begin{split}
    N_{\rm exp}&=\sum\limits_j N_j T\dfrac{\rho_\chi}{m_\chi}\int dv_f ~v_f f(v_f)~\\
    &\,\times\int dE_R~\dfrac{d\sigma_j}{dE_R}\int dE\, Res(E,E_R)\epsilon(E)\,,
\end{split}
\label{eq:Nexp}
\end{equation}
where $N_j$ is the number of target nuclei $j$ in the detector, $T$ is the duration of data taking, $\epsilon(E)$ is the detection efficiency, and $Res(E,E_R)=\frac{1}{\sqrt{2\pi}\sigma_d}\exp(-(E-E_R)^2/(2\sigma_d^2))$ is a Gaussian detector response with energy resolution $\sigma_d$. 
 The density of DM near the Earth is $\rho_\chi=0.3$~GeV/cm$^3$ and its velocity distribution upon arriving at the detector is $f(v_f)$. If dark matter experiences strong attenuation in the overburden, $f(v_f)$ may deviate significantly from the Maxwell distribution in the halo.

Since we focus on dark matter  with $m_\chi < 10~\gev$, for which the momentum transfer in 
each scatter is very small, we set the form factor $F_j(E_R)=1$. 
In this case, the nuclear scattering cross section is
\begin{equation}
    \sigma_{j}\simeq \dfrac{\mu^2_{A_j}}{\mu_N^2}\fiv^2 \sigma_0\,,
\end{equation}
where $\mu_{A_j}$ is the reduced mass of the $\chi$-nucleus system.  If the cross section is small enough that $\chi$ rarely scatters before reaching the detector, the number of events is directly proportional to $\sigma_j$. Then, the number of events at the detector for IVDM is scaled by the corresponding number for isospin-conserving interactions ($f_p=f_n=1$) by a {\it detector factor},
\begin{equation}
    R_D=\dfrac{\sum_j N_j \mu^2_{A_j}\fiv^2}{\sum_j N_j \mu^2_{A_j}A_j^2}\,.
    \label{eq:Rsigma}
\end{equation}

\begin{figure*}[t]
    \includegraphics[trim={0.3cm 0cm 0.3cm 0cm},clip,width=0.495\textwidth]{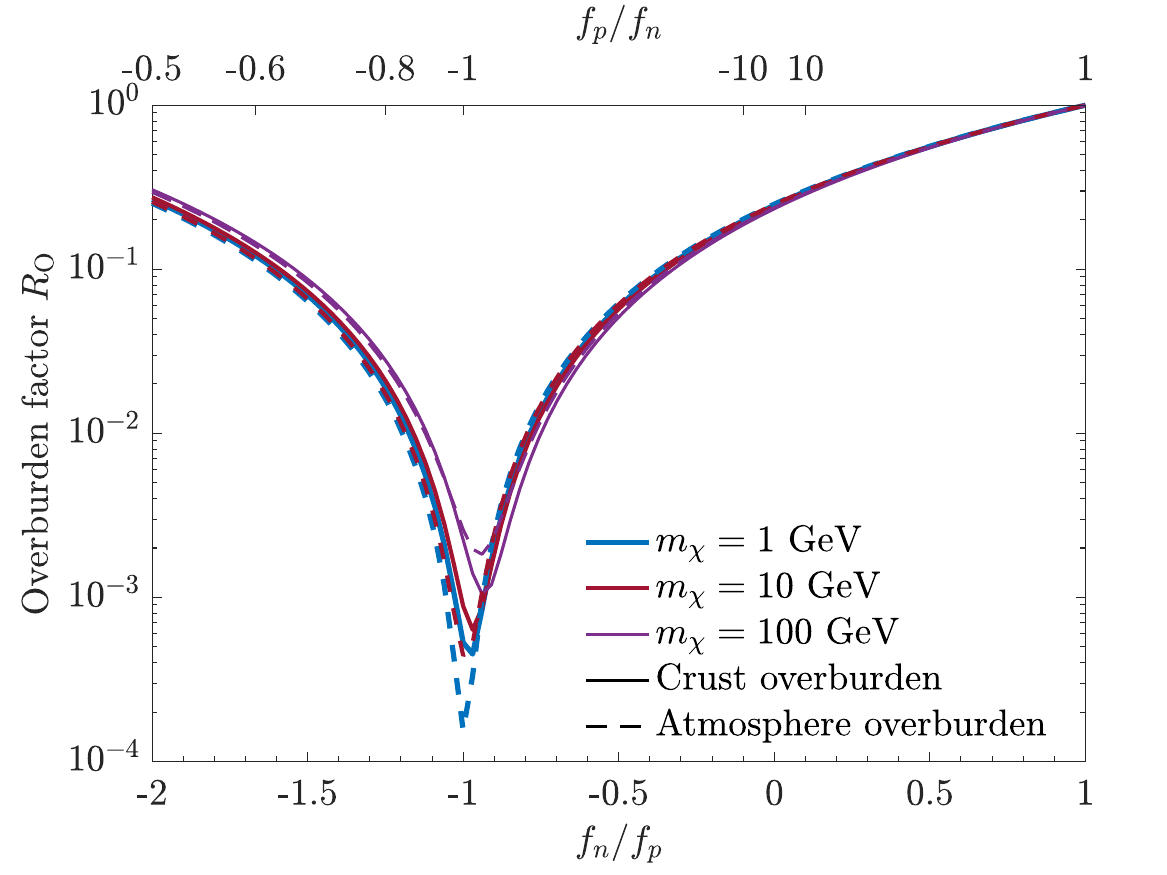}
    \includegraphics[trim={0.3cm 0cm 0.3cm 0cm},clip,width=0.495\textwidth]{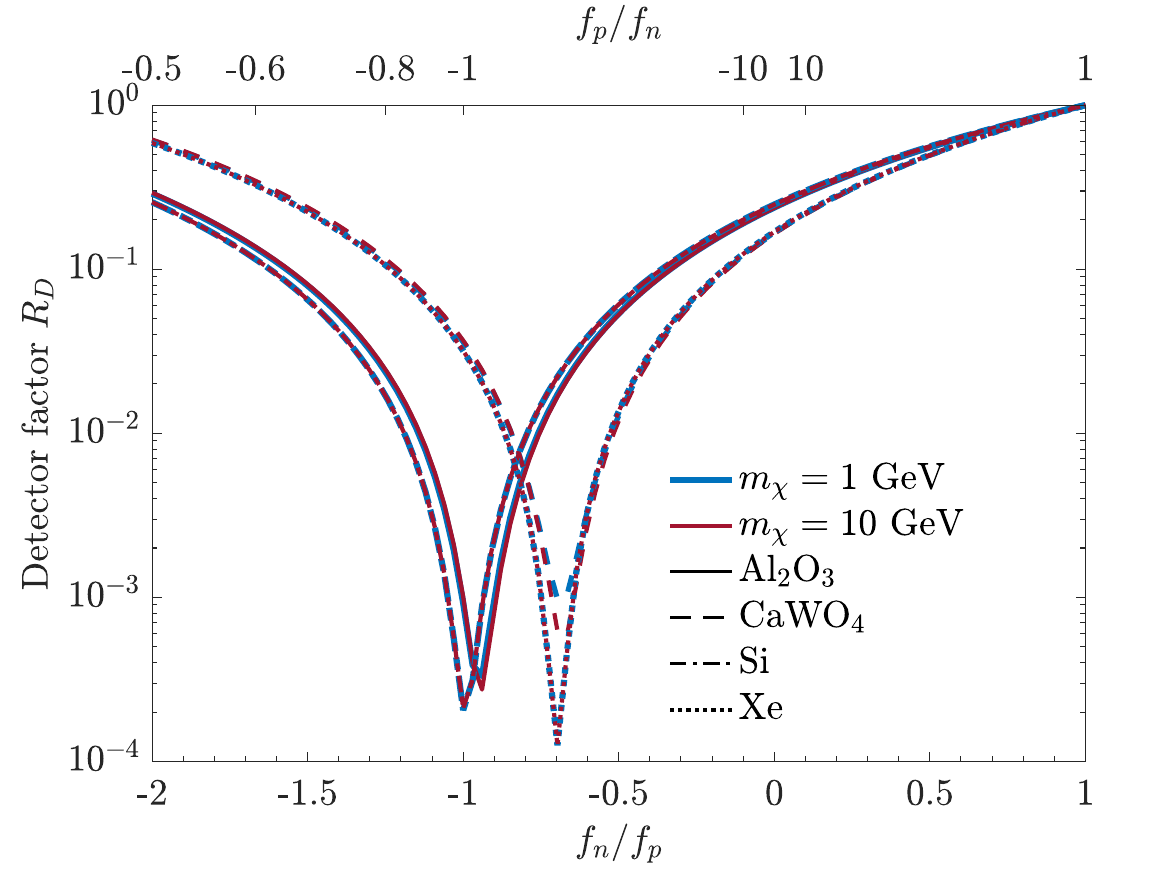}
    \caption{Effects of isospin-violating interactions in direct detection experiments. {\it Left:} Overburden factor as defined in Eq.~\eqref{eq:ROF} for three DM masses. The solid (dashed) curves correspond to the crust (atmosphere) as the overburden. {\it Right:} The reduction in the DM event rate given by the detector factor in Eq.~\eqref{eq:Rsigma}. The solid, dashed, dash-dotted and dotted lines correspond to Al$_2$O$_3$, CaWO$_4$, Si and Xe as the detector material.} 
    \label{fig:Rfactors}
\end{figure*}

In the opposite limit of a large enough cross section, $\chi$ may scatter multiple times before reaching the detector. The DM velocity is reduced in each scattering, with the final velocity exponentially dependent on the scattering cross section. To see this quantitatively, we assume $\chi$ travels along a straight line, so that the change in its velocity per unit distance can be expressed as~\cite{Kavanagh:2017cru}
\begin{equation}
    \dfrac{dv}{dx}=-v\Phi\,,\ \ \Phi\equiv \sum_j\limits\dfrac{\mu^2_{A_j}n_j}{m_\chi m_{A_j}}\sigma_j\,,
\end{equation}
where $n_j$ is the number density of nucleus $j$ in the overburden. 
Starting from an initial velocity $v_i$, the final velocity after traversing the overburden is $v_f=v_i\exp(-\int \Phi~dx)$
and the DM velocity distribution at the detector is
\begin{equation}
    f(v_f)=f_h (v_i)\exp\left(\int \Phi~ dx\right)\,.
    \label{eq:fvf}
\end{equation}
Here, $f_h (v_i)$ is the halo DM velocity distribution in the frame of the Earth. We take the Earth's velocity with respect to the galactic rest frame $v_\oplus=232$~km/s and the escape velocity from our galaxy $v_{\rm esc}=544$~km/s. The number of events in the detector is therefore sensitive to the exponent $\Phi$, which is in turn affected by isospin-violating effects. To quantify how large this effect is, we define the {\it overburden factor},
\begin{equation}
    R_O=\dfrac{\Phi(f_p,f_n)}{\Phi(1,1)}=\dfrac{\sum_j n_j \mu^4_{A_j}\fiv^2/m_{A_j}}{\sum_j n_j \mu^4_{A_j}A_j^2/m_{A_j}}\,.
\label{eq:ROF}
\end{equation}
This is the factor by which $\Phi$ is scaled to account for isospin violation. A smaller value of 
$R_O$ leads to a velocity distribution 
at the detector which is more similar to $f_h (v_i)$.
We use the PREM~\cite{dziewonski1981preliminary} model for the Earth composition and the NRLMSISE-00 atmosphere model~\cite{NRLMSISE-00} with average oxygen, nitrogen, argon mass fractions of 23.18\%, 75.6\%, and 1.2\% respectively.

The isospin-violating effects parameterized in terms of $R_O$ and $R_D$ are displayed in Fig.~\ref{fig:Rfactors}. In the left panel we show overburden effects 
in the crust and in the atmosphere separately, as in most direct detection experiments only one of the two dominates. In this way the fractional abundance of the chemical elements stays approximately constant in the overburden. Since most of the chemical elements in the Earth have nearly equal number of protons and neutrons, $R_O$ has a minimum close to $f_n=-f_p$. 
If $m_\chi \lesssim \gev$, then $\mu_{A_j} \sim m_\chi$ and $R_O$ becomes independent of $m_\chi$.
For heavier DM, $\mu_{A_j} \sim m_\chi (1 - m_\chi / m_{A_j})$, and for heavier (neutron-rich) nuclei $f_{\rm IV}$ is reduced for $|f_n/ f_p| <1$. These two facts taken together imply that for heavier DM, heavier nuclei affect the numerator of 
$R_{O}$ more than the denominator.
This also explains why the minimum of $R_O$ moves to a slightly smaller value of  $|f_n/f_p|$ as 
$m_\chi$ increases. In the right panel we show the reduction in the expected number of events in the detector for various detector materials considered in Section~\ref{sec:IVDMExp}. The location of the trough is mainly determined by the dominant scattering target in the detector. For CaWO$_4$ and Xe, the trough $f_n/f_p$ ratio is close to $-0.7$. For other materials the ratio is close to $-1$. Note that both $R_O$ and $R_D \gtrsim 10^{-4}$. 

\section{Velocity distribution: three approaches}
\label{sec:IVDMComp}

 We use three approaches to quantify the effect of $\chi$ scattering in the overburden on the DM velocity distribution at the detector: the straight-line approximation, an analytical approach, and full Monte Carlo simulations.

\textit{\textbf{Straight line approximation (SLA).}}
In this approximation, the DM velocity distribution is given by Eq.~\eqref{eq:fvf}. We assume that $\chi$ does not change its direction markedly during propagation and the particles arrive from an average zenith angle of $54^\circ$~\cite{Hooper:2018bfw}. The effect of scattering is  only to degrade the DM velocity, while leaving the total DM flux unchanged. SLA works well for heavy DM whose scattering is more forward-peaked. However, for light DM, the scattering is more isotropic, with a fraction of DM particles reflected away from the Earth and never reaching the detector~\cite{Bramante:2022pmn}.

\begin{figure}[t]
    \centering
    \includegraphics[width=\columnwidth]{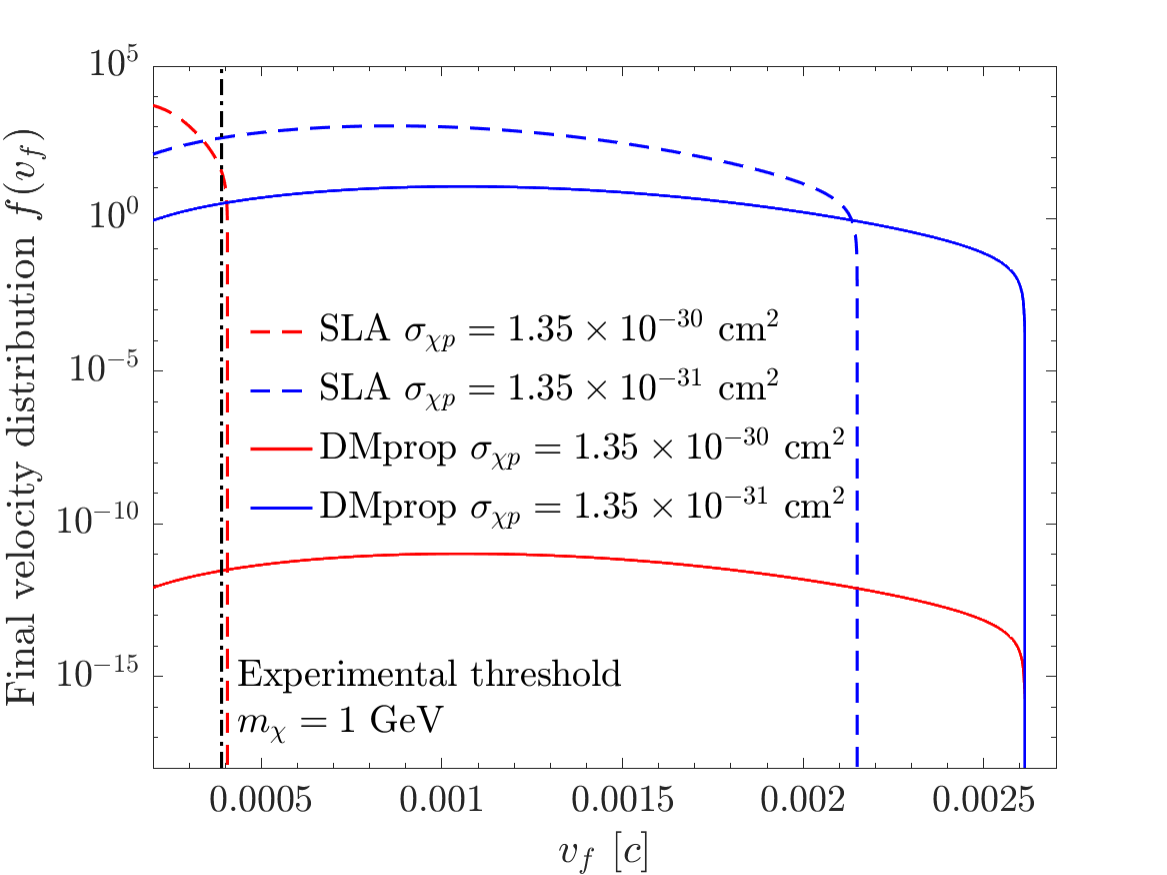}
    \caption{Velocity distribution of isospin-conserving dark matter after traversing the overburden to a detector 1400~m underground. Dashed lines use the straight line approximation (SLA) and solid lines use the analytical calculation as implemented in \texttt{DMprop}~\cite{Cappiello:2023hza}. 
    The vertical dot-dashed line marks the minimum threshold  velocity for a 1~GeV mass $\chi$ at CRESST-III~\cite{CRESST:2019jnq}. }
    \label{fig:veldist}
\end{figure}

Examples of dark matter velocity distributions at the detector using SLA are shown in Fig.~\ref{fig:veldist}. As the scattering cross section is increased, the maximum DM velocity falls, and the distribution peaks at lower velocities. The overall normalization remains unchanged.

\textit{\textbf{Analytical approach.}} The analytical treatment of dark matter propagation is discussed in Ref.~\cite{Cappiello:2023hza}. 
Assuming a constant matter density in the overburden, the probability that $\chi$ reaches the detector after scattering $M$ times can be evaluated recursively via
\begin{equation}
    P_M(z)=\int_0^\infty P_{M-1}(z')\dfrac{1}{2l}\Gamma(0,|z-z'|/l)dz'\,,
\end{equation}
where $z$ is the depth of the detector, $l$ is the mean free path of $\chi$ in the overburden and $\Gamma$ is the incomplete gamma function. The probability of not scattering is $P_0(z)=-l^{-1}Ei(-z/l)$ where $Ei$ is the exponential integral. The DM energy spectrum after $M$ scatters, $dN/dE_M$, can also be computed recursively by integrating over the energy loss (or scattering angle) distribution, assuming that the scattering is isotropic in the frame of the Earth. The DM velocity distribution at the detector is
\begin{equation}
    f(v_f)=\dfrac{m_\chi v_f}{N_0}\dfrac{dN}{dE}=\dfrac{m_\chi v_f}{N_0}\sum\limits_{M=0}^\infty P_N(z)\dfrac{dN}{dE_M}(v_f)\,,
\end{equation}
where $dN/dE_M$ is the DM flux (at the detector) of particles which have scattered $M$ times, 
$dN/dE$ is the total DM flux obtained after summing over $M$, and $N_0$ is a normalization factor. As both downward and upward scattering is accounted for, the DM flux for $m_\chi\lesssim m_{A_j}$ is reduced due to reflection, which is more pronounced for large $M$.

Examples of velocity distributions at the detector computed using the analytical approach as implemented in the \texttt{DMprop}~\cite{Cappiello:2023hza} code are presented in Fig.~\ref{fig:veldist}. We sum up the probability of reaching the detector after up to 100 scatters, beyond which the DM flux becomes negligibly small and only contributes to the very low velocity tail of the distribution. As the cross section increases, the flux of particles that can reach the detector diminishes dramatically. However, because the probability that the DM particles merely undergo a few scatters before reaching the detector is nonzero, the high velocity part of the distribution is still appreciable even for large cross sections.
This is in contrast with the SLA approach, which assumes that every particle scatters the expected number of times, and experiences the expected energy loss in each scatter.

\textbf{\textit{Monte Carlo (MC).}} A third approach is to simulate the trajectories of DM particles inside the Earth. 
The angle between the position vector of the detector (with
the origin at the Earth's center) and $\vec{v}_\oplus$ (taken as the z-axis) is the {\it isodetection angle} $\Theta$, which by virtue of azimuthal symmetry, defines a ring of constant flux called an {\it isodetection ring}~\cite{Collar:1992qc}. 
The DM velocity distribution at the detector is computed by evaluating the weighted sum of trajectories that traverse the isodetection ring, and inherits the azimuthal symmetry.
We employ the \texttt{DaMaSCUS} code~\cite{Emken:2017qmp}, modified to accommodate isospin-violating interactions. Since the code normalizes the integral of the velocity distribution to unity, the reduction of DM flux due to reflection is not taken into account. In general, we expect reflection to suppress the normalization of the velocity distribution at large isodetection angles more than at small angles as the DM particle undergoes more scattering before it reaches the detector.
If $\chi$ is much lighter than the nuclei it scatters against, then the fraction which remains in Earth after $M$ scatters 
scales roughly as $M^{-1/2}$~\cite{Neufeld:2018slx}.
 The extension of the code to include this effect is deferred to future work.

 \begin{figure*}[!htb]
    \includegraphics[trim={0.25cm 0cm 0.25cm 0cm},clip,width=0.495\textwidth]{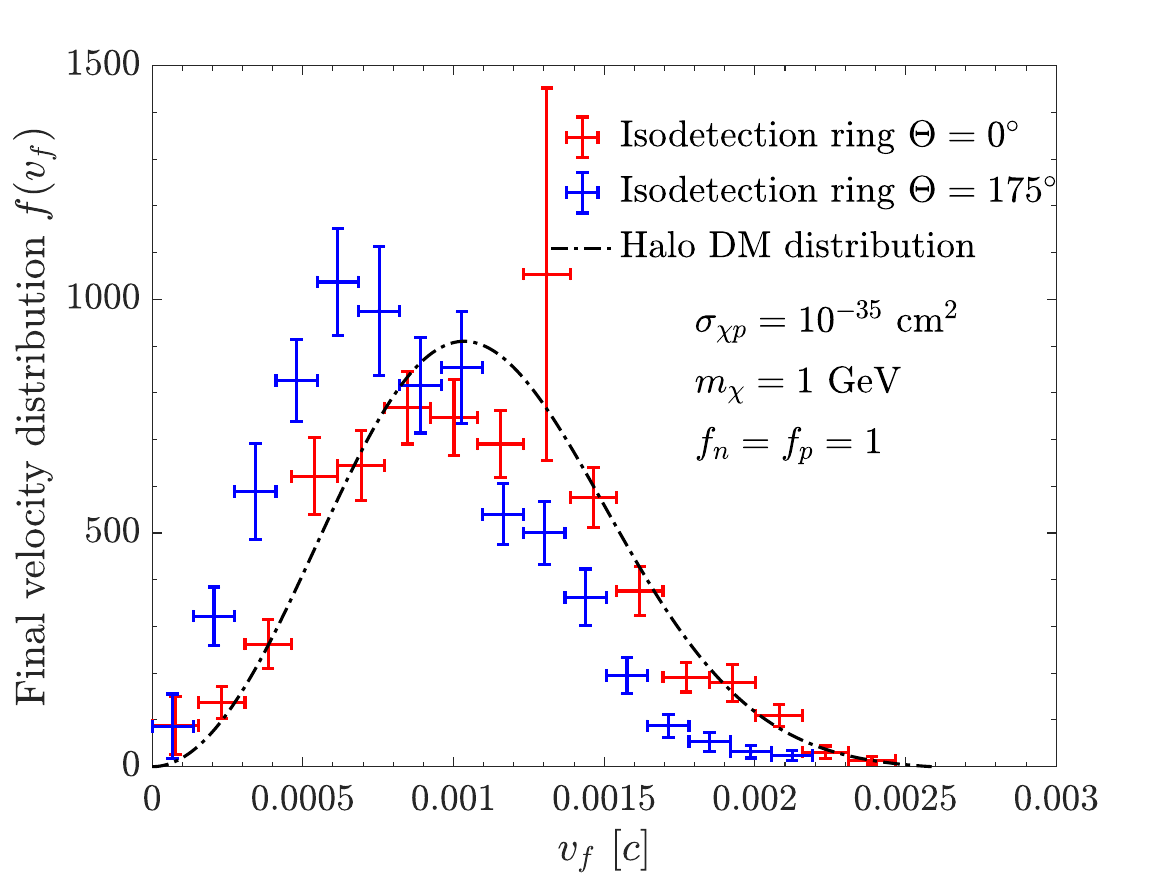}
    \includegraphics[trim={0.25cm 0cm 0.25cm 0cm},clip,width=0.495\textwidth]{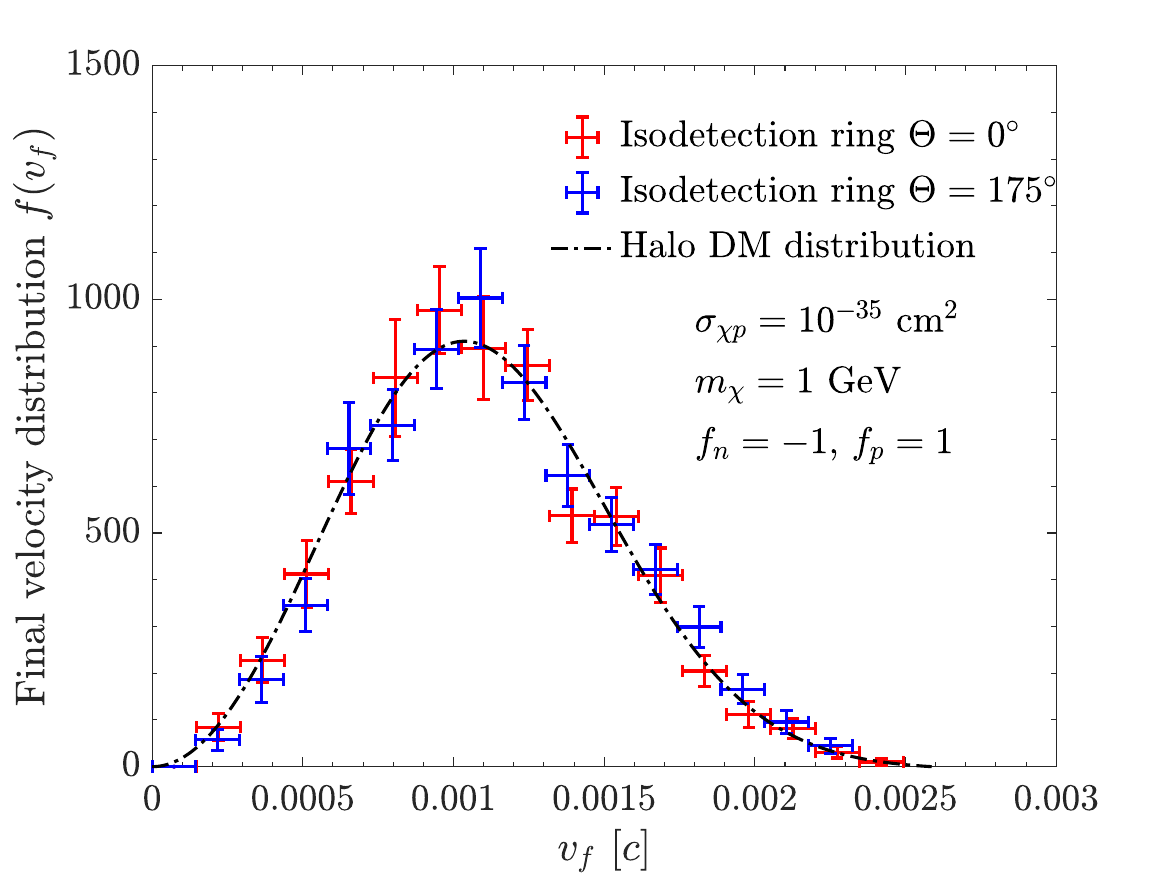}
    \caption{DM velocity distributions at two isodetection angles for a direct detection experiment located 1400~m underground using \texttt{DaMaSCUS}.
    The dot-dashed curve is the halo DM distribution in the Earth's frame.
    We assume isospin-conserving interactions in the left panel, and isospin-violating interactions with $f_p=-f_n=1$ in the right panel.
    The dot-dashed curve is the halo DM distribution in the Earth's frame. We set $m_\chi=1$~GeV and $\sigma_{\chi p}=10^{-35}$~cm$^2$. }
    \label{fig:veldistDaM}
\end{figure*}

Examples of DM velocity distributions from the MC are displayed in Fig.~\ref{fig:veldistDaM}. The DM flux is binned to estimate the distributions and the corresponding uncertainties. For $\sigma_{\chi p}=10^{-35}$~cm$^2$, a shift in the velocity distribution to lower values is evident for isospin-conserving interactions for $\Theta=175^\circ$ (i.e., DM mainly arrives from the opposite side of the Earth), compared with $\Theta=0^\circ$ (i.e., DM mainly arrives from above the detector). However, for isospin-violating interactions with $f_p=-f_n=1$, the distributions match the halo DM distribution at both isodetection angles because the overburden effect is reduced. A Monte Carlo simulation is ideal for studying the daily modulation of dark matter since a different overburden is expected at different times due to the changing orientation of the DM flux with respect to the detector. Nonetheless, we do not use the full MC to set constraints on the DM scattering cross section because the DM flux falls quickly as the cross section increases, and the number of DM particles required to establish a reliable velocity distribution becomes computationally prohibitive. MC simulations that use planar layers instead of a spherical Earth can be performed using the \texttt{DaMaSCUS-CRUST} code~\cite{DaMaSCUScrust,Emken:2018run,Emken:2019tni}. 
The analytic approach generally shows good agreement with \texttt{DaMaSCUS-CRUST}, but is much more computationally 
efficient, especially for large cross sections~\cite{Cappiello:2023hza}.

\section{Experimental Constraints}
\label{sec:IVDMExp}

\begin{figure*}[!htb]
\includegraphics[trim={0cm 0cm 0.3cm 0cm},clip,width=0.495\textwidth]{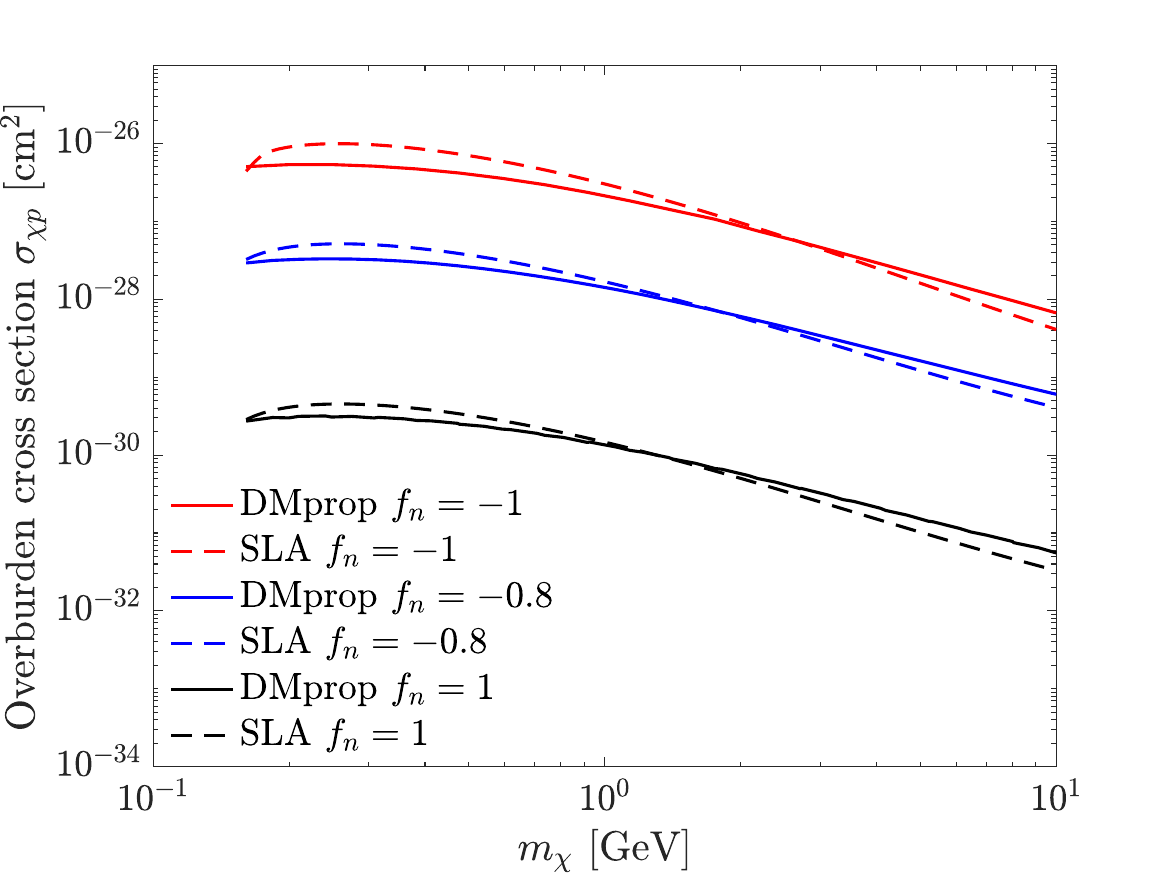}
    \includegraphics[trim={0cm 0cm 0.3cm 0cm},clip,width=0.495\textwidth]{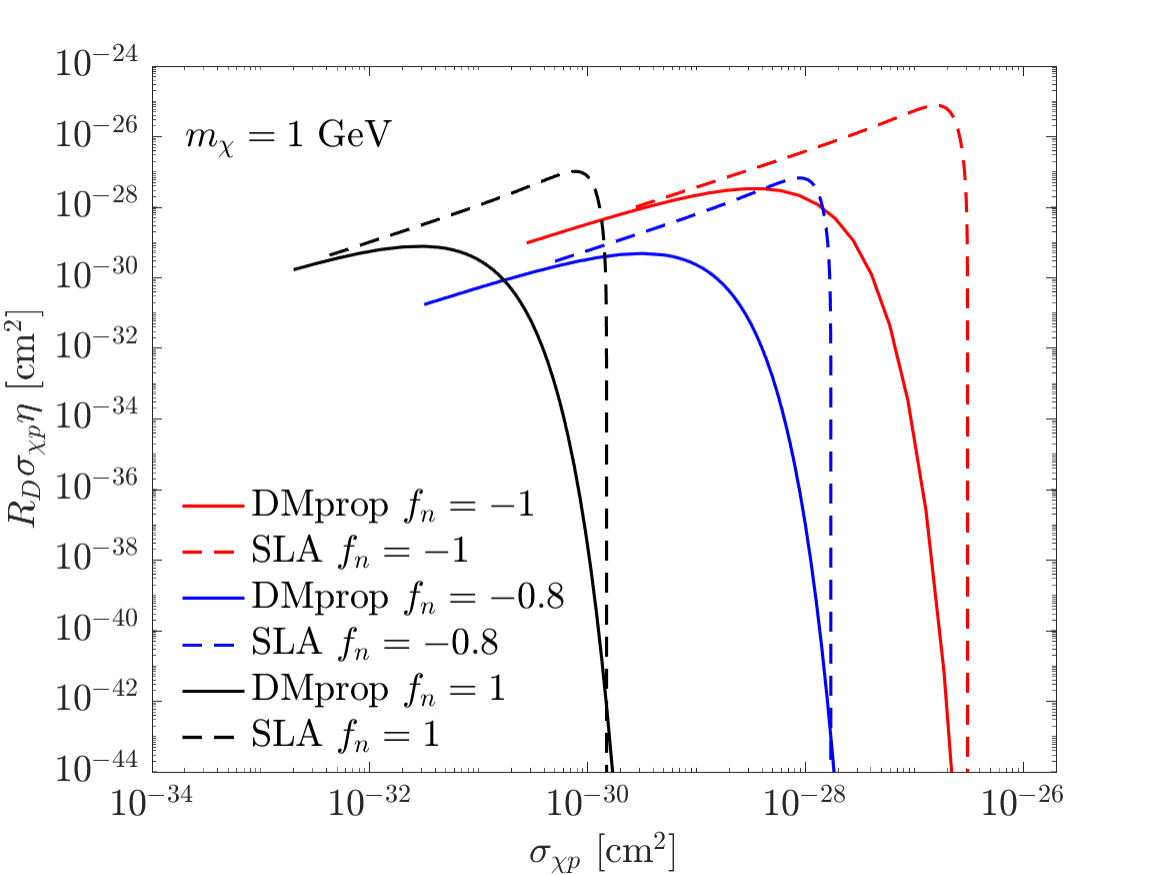}
    \caption{Comparison between straight line approximation (SLA) and the analytical approach using the \texttt{DMprop} code. The dashed lines correspond to SLA and the solid lines are obtained from \texttt{DMprop} for various values of $f_n$ and $f_p=1$. 
    \textit{Left:} The overburden cross section for the CRESST-III~\cite{CRESST:2019jnq} experiment located 1400~m underground.
    \textit{Right:} A measure of the number of detectable events at CRESST-III; see Eq.~\eqref{eq:Nexpeta}.}
    \label{fig:upperlimits}
\end{figure*}

To determine the region of $(m_\chi, \sigma)$ parameter space excluded by a direct detection experiment, we only need the upper and lower boundaries 
of the excluded region for any $m_\chi$. Since this does not depend on the details of the velocity distribution at the detector within the interior of the excluded region, 
we find that the analytical approach is sufficient.  In particular, at the lower boundary of the excluded region, the effect of the overburden is negligible, and any of three 
approaches is satisfactory.  But at the upper boundary, scattering in the overburden dramatically reduces the number of $\chi$ that 
reach the detector with a velocity large enough to produce a signal above threshold.  In this case, the dominant contribution to the experimental event rate comes from particles 
arriving from nearly overhead, with the smallest path lengths through the overburden. For these particles, the analytical approach is expected to be adequate.  Moreover, the assumption of isotropic scattering in the frame of the 
Earth is a good approximation for light DM.  Although the analytical approximation is expected to be 
better than SLA for light DM, it is interesting to see how much these approaches can deviate from each other.

We call the maximum scattering cross section that can be constrained by an experiment, the {\it overburden cross section}. For CRESST-III, this is shown in the left panel of Fig.~\ref{fig:upperlimits}. Generally, the results of SLA agree with the analytical approach. 
To understand this behavior, we show the integral of the relevant distribution in the right panel of Fig.~\ref{fig:upperlimits}.  We  define
\begin{equation}
    \eta (E_R) =\int_{v_{f,\min}}^{v_{f,\max}}f(v_f)/v_f~dv_f\,,
    \label{eq:eta}
\end{equation}
where $v_{f,\min}$ is the minimum DM speed at the detector to produce 
an elastic recoil with energy $E_R$, and $v_{f,\max} \simeq v_\oplus+v_{\rm esc}$.
Note, $\eta$ depends implicitly on $f_n / f_p$, $\sigma_{\chi p}$, and the depth of the detector.
The expected number of events in Eq.~\ref{eq:Nexp} can then be written as
\begin{equation}
\begin{split}
    N_{\rm exp}&=\sum\limits_j N_j T\dfrac{\rho_\chi}{m_\chi}\dfrac{m_{A_j}\sigma_0\fiv^2}{2\mu_N^2}\\
    &\times\int\,\eta\,dE_R\int\, Res(E,E_R) \epsilon(E) \,dE\,.
\end{split}
\end{equation}
For $m_\chi=1$~GeV and for a detector with an energy threshold of CRESST-III, we may approximate 
\begin{eqnarray}
N_{\rm exp}   &\sim & \sum\limits_j N_j T\dfrac{\rho_\chi}{m_\chi}\dfrac{m_{A_j} A_j^2 }{2\mu_N^2} 
\left[ \int\epsilon\, dE_R \right] 
\nonumber\\
&\,& 
\times R_D \sigma_{\chi p} \eta\,,
\label{eq:Nexpeta}
\end{eqnarray}
by evaluating $\eta$ with $v_{f,\min} = 100$~km/s.  We plot $R_D \sigma_{\chi p} \eta$ as a function of 
$\sigma_{\chi p}$ in the right panel of Fig.~\ref{fig:upperlimits} for various values of $f_n / f_p$. 

Because $R_D \sigma_{\chi p} \eta$ is proportional to the number of  scattering events above the 
experimental threshold, any choice of $m_\chi$, detector material, geometry, and background rate, will yield a 
minimum value of $R_D \sigma_{\chi p} \eta$ for which a choice of $f_n/f_p$  can be excluded by a null detection.\footnote{For example, 
for $m_\chi = 1~\gev$, and for a detector with the properties of CRESST-III, values of $f_n/f_p$ can be excluded by a null detection if $R_D \sigma_{\chi p} \eta \gtrsim 10^{-42}~\cm^2$. } 
In general, the 
range of $\sigma_{\chi p}$ for which a set of IVDM parameters is excluded is bounded from above and below.  For small $\sigma_{\chi p}$, 
the expected number of events grows linearly with $\sigma_{\chi p}$, regardless of $f_n / f_p$.  This is expected in the limit of small scattering cross section, since the effect of the overburden is 
negligible.  In the SLA approach, this behavior continues until scattering in the overburden becomes significant 
enough that the flux of DM particles able to produce detectable scattering 
events is attenuated. Under the assumption that each particle scatters the expected 
number of times and loses the expected amount of energy per scatter,
the number of detectable DM particles 
vanishes and the event rate drops sharply to zero.

This behavior is smoothed out in the analytical approach by reflection and statistical fluctuations 
in the number of scatters and energy loss per scatter.  In particular, for intermediate values of $\sigma_{\chi p}$, 
the flux of particles capable of producing detectable scatters is reduced, relative to SLA, because of 
reflection. And for large $\sigma_{\chi p}$, for which the flux of detectable particles vanishes in SLA, a 
flux of detectable particles is still present in the analytical approach, due to the tail of particles which have scattered 
fewer times than expected in the overburden.  

The difference between SLA and the analytical approach is illustrated in Fig.~\ref{fig:veldist}. From the left panel of Fig.~\ref{fig:upperlimits}, the overburden cross section is about $1.35\times 10^{-30}$~cm$^2$ for $m_\chi = 1~\gev$ and $f_n/f_p=1$. For such a large cross section, in SLA, the maximum velocity of DM particles at the detector is only slightly above threshold, which substantially suppresses the fraction of $\chi$ that can trigger the detector. In the analytical approach, the DM flux is degraded so that only a minuscule part of it can reach the detector. The different treatment of reflections and statistical fluctuations in the number of scatters in the overburden, leads to a smoother transition from small to large cross sections in the analytical approach than with SLA, and the two curves for a given value of $f_n$ in the right panel of Fig.~\ref{fig:upperlimits} eventually intersect at some cross section. The reflection effect also accounts for the less pronounced peak in the analytical approach.

However, despite the difference, once the cross section is large enough for the overburden to become important, the reduction of $v_{f,\max}$ in SLA and the reduction of the DM flux in the analytical approach are both driven by the fact that multiple scattering takes place in the overburden, i.e., the mean free path of DM is much smaller than the depth of the detector. Consequently, the overburden cross sections computed in these two approaches coincide with each other, with discrepancies attributable to the different velocity distributions. 
Essentially, although SLA does not correctly reproduce the velocity-distribution at the detector, it 
does roughly determine the cross section above which particles at the detector are too slow to produce detectable scatters.  For the 
overburden cross section, particles have scattered so often in the overburden that very few particles reaching the detector can 
produce detectable scatters, although a relatively large fraction of those are detected.  So the overburden cross section 
is close to the cross section at which the event rate drops to zero in SLA. The overburden cross sections obtained for CREST-III with the two approaches match because the curves in the right panel of Fig.~\ref{fig:upperlimits} fall very steeply.\footnote{However,
if the exclusion of a set of IVDM parameters requires a large value of $R_D \sigma_{\chi p} \eta$ because of a different experimental setup (e.g., if the detector has a small target mass or large background rate), then that value of $R_D \sigma_{\chi p} \eta$ will
 intersect the SLA and \texttt{DMprop} curves at different points in the right panel of Fig.~\ref{fig:upperlimits}, resulting in 
a large difference in the overburden cross sections obtained with the two approaches.}

As $f_n/f_p$ approaches $-1$, the overburden cross section curves shift to smaller values. This behavior is qualitatively understood from Fig.~\ref{fig:Rfactors}, because $R_O$ has a minimum near $f_n/f_p\simeq -1$. The velocity integral $\eta$ also increases for smaller $f_n/f_p$, which effectively reduces the overburden. However, the maximum number of events in the detector, indicated by $R_D \sigma_{\chi p} \eta$, does drop slightly for $f_n/f_p=-0.8$, due to the trough in $R_D$ near $f_n/f_p\sim -0.7$.

Although the  
analytical approach and SLA yield similar results, we adopt the former in our calculations because it is better motivated physically.
 The assumption of isotropic scattering in \texttt{DMprop} is only valid for $m_\chi\ll m_A$, so we truncate $m_\chi$ at 10~GeV. We derive constraints from the following direct detection experiments by requiring the expected number of events in Eq.~\eqref{eq:Nexp} to not exceed the detected number of DM candidates at 90\% CL.

\begin{figure*}[!htb]
    \centering
    \includegraphics[trim={0cm 0cm 0.3cm 0cm},clip,width=0.495\textwidth]{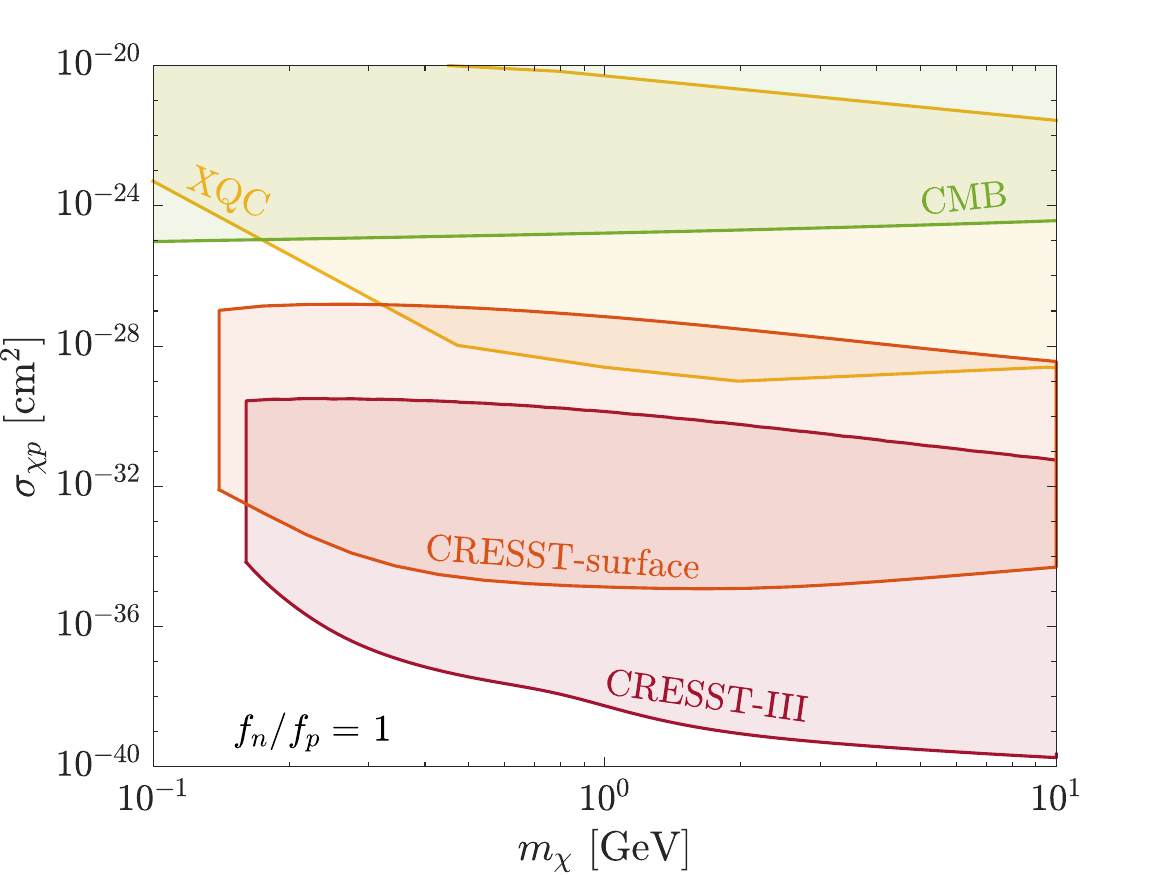}
    \includegraphics[trim={0cm 0cm 0.3cm 0cm},clip,width=0.495\textwidth]{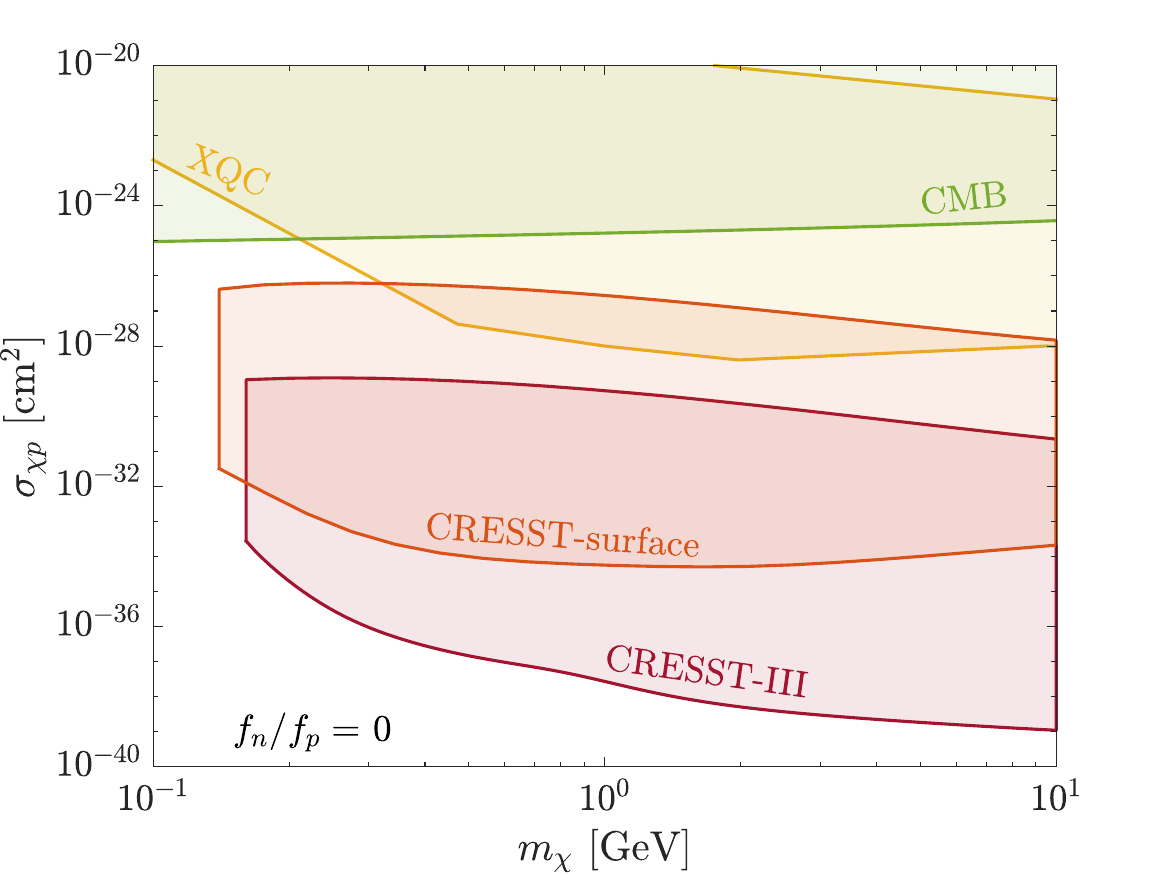}
    \includegraphics[trim={0cm 0cm 0.3cm 0cm},clip,width=0.495\textwidth]{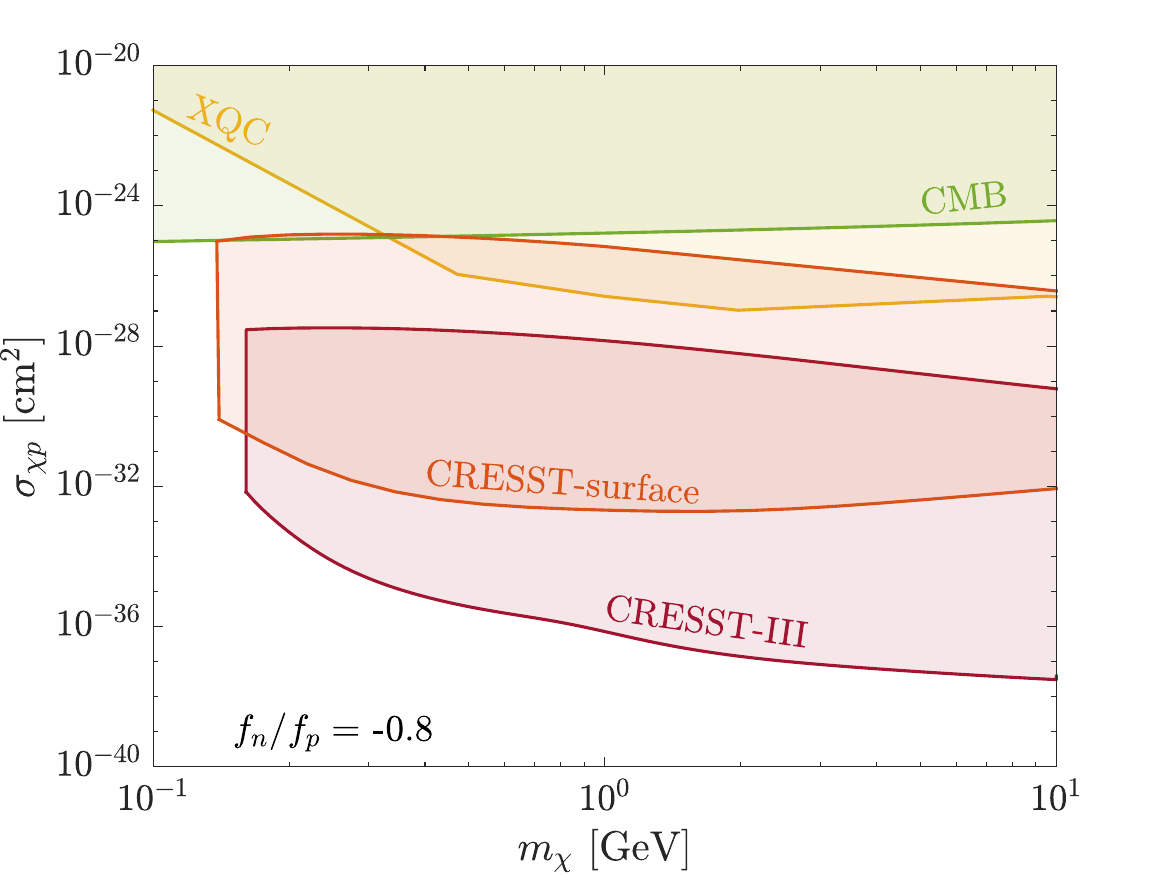}
    \includegraphics[trim={0cm 0cm 0.3cm 0cm},clip,width=0.495\textwidth]{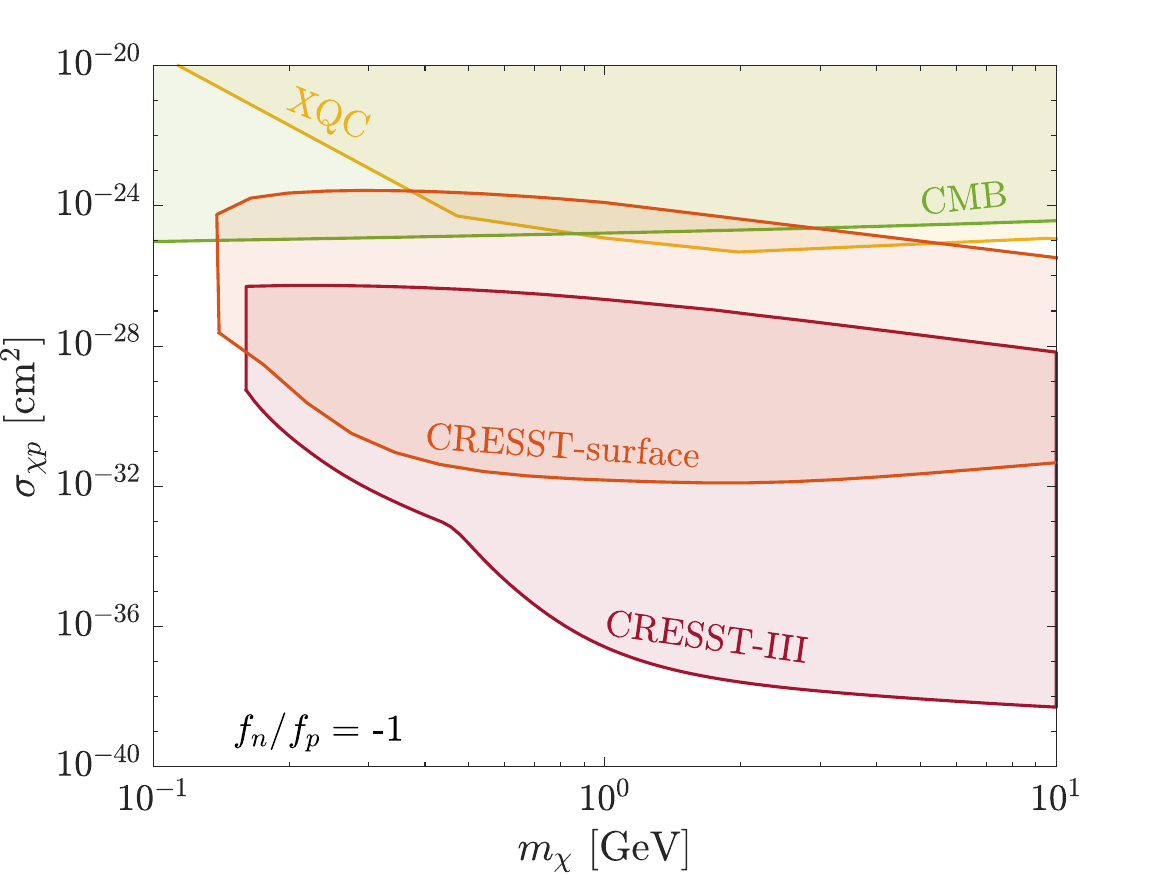}    
    \caption{90\% CL bounds on the dark matter-proton scattering cross section $\sigma_{\chi p}$ for several values of $f_n/f_p$ with $f_p=1$.  We also show the 95\% CL bound from the CMB~\cite{Dvorkin:2013cea,Gluscevic:2017ywp}. The 90\% CL region excluded by XQC is obtained by rescaling the result of Ref.~\cite{Erickcek:2007jv}.}
    \label{fig:results_fntofp}
\end{figure*}

\begin{figure*}[!htb]
    \centering
    \includegraphics[trim={0cm 0cm 0.3cm 0cm},clip,width=0.495\textwidth]{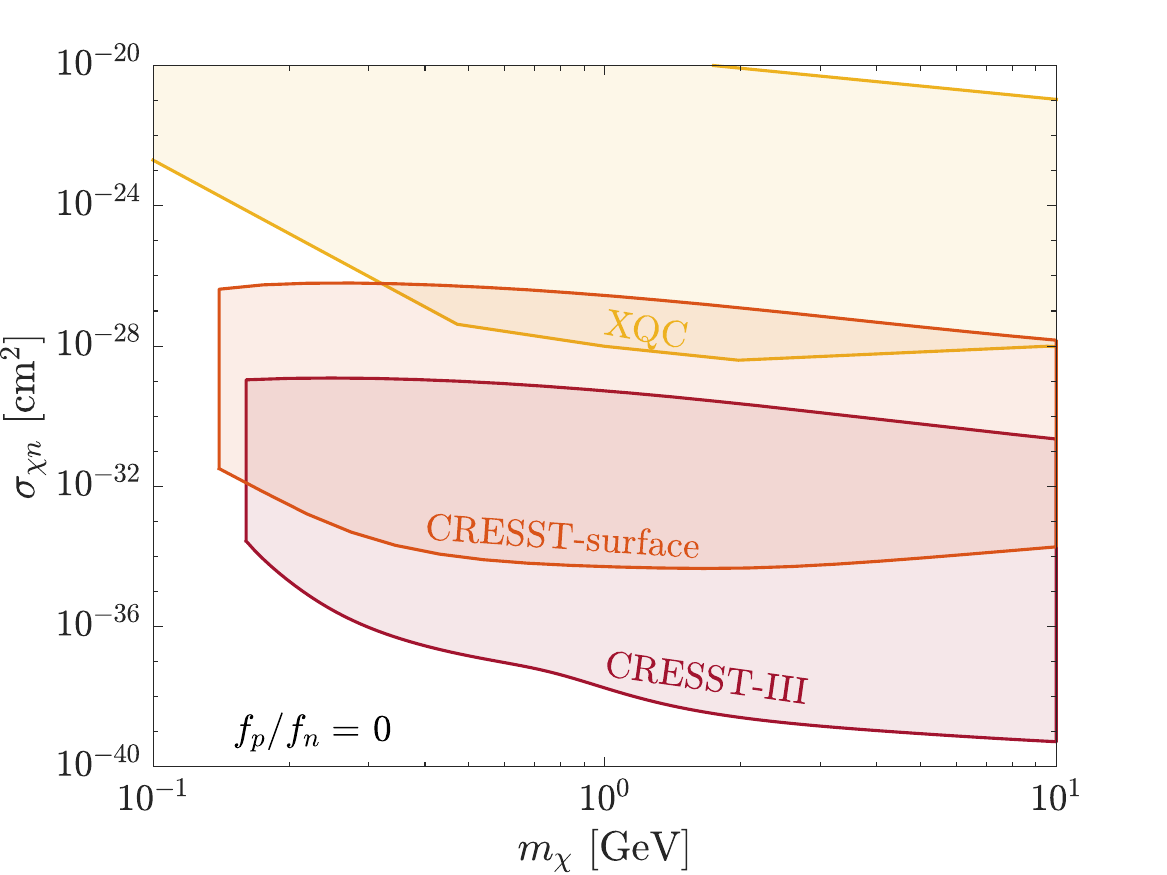}
    \includegraphics[trim={0cm 0cm 0.3cm 0cm},clip,width=0.495\textwidth]{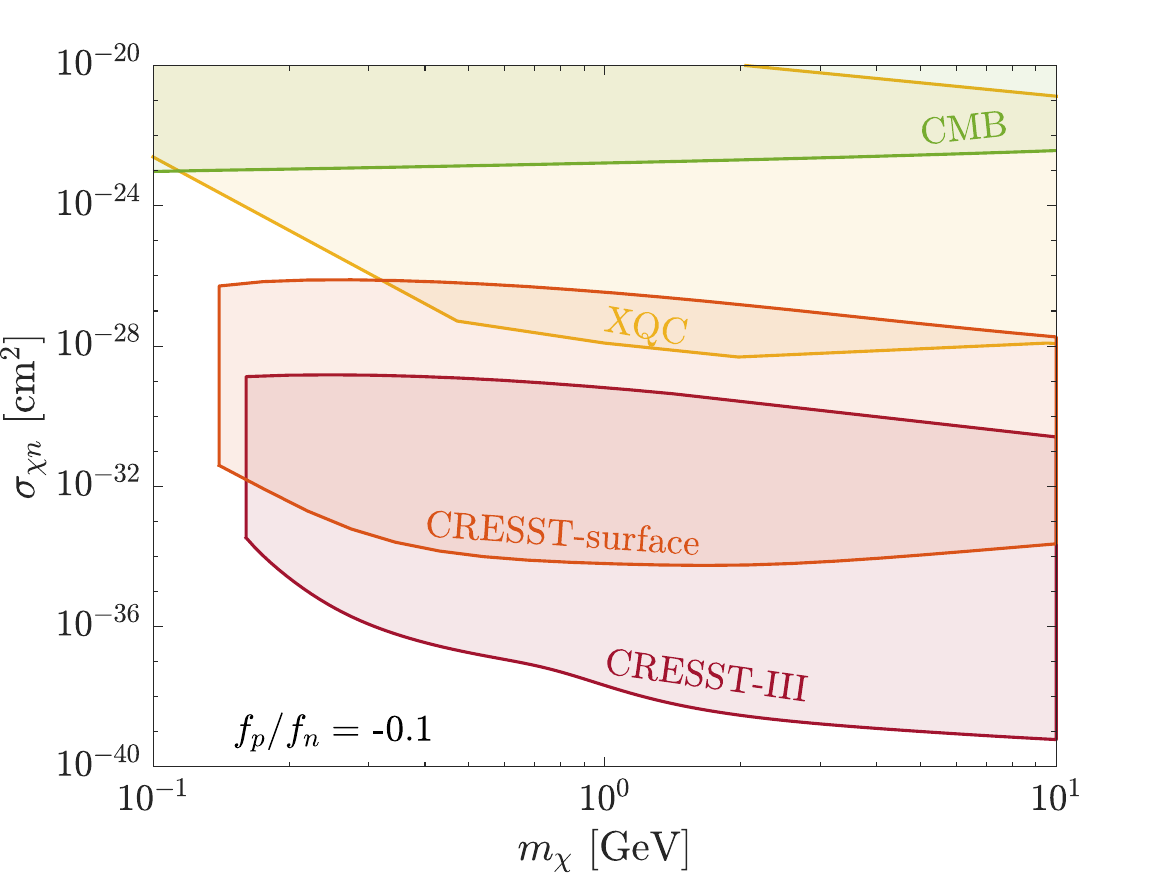}
    \includegraphics[trim={0cm 0cm 0.3cm 0cm},clip,width=0.495\textwidth]{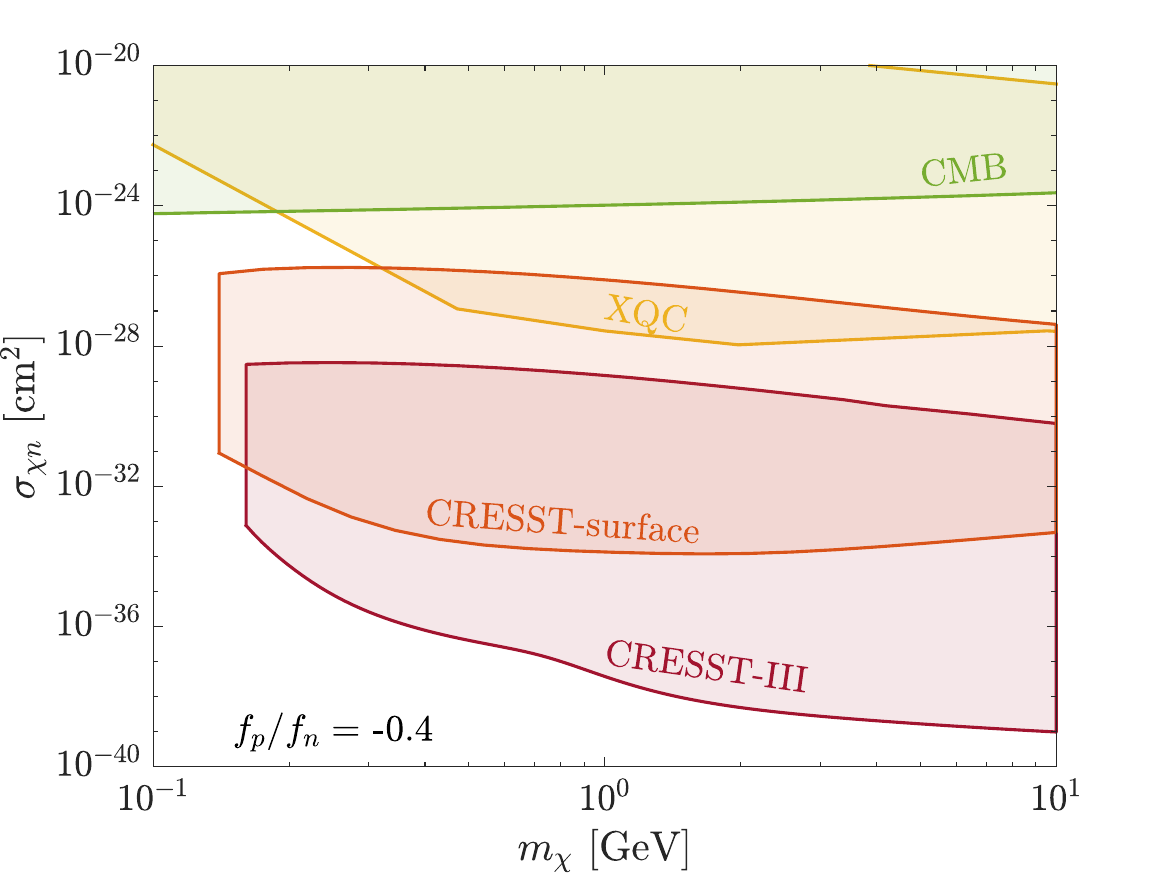}
    \includegraphics[trim={0cm 0cm 0.3cm 0cm},clip,width=0.495\textwidth]{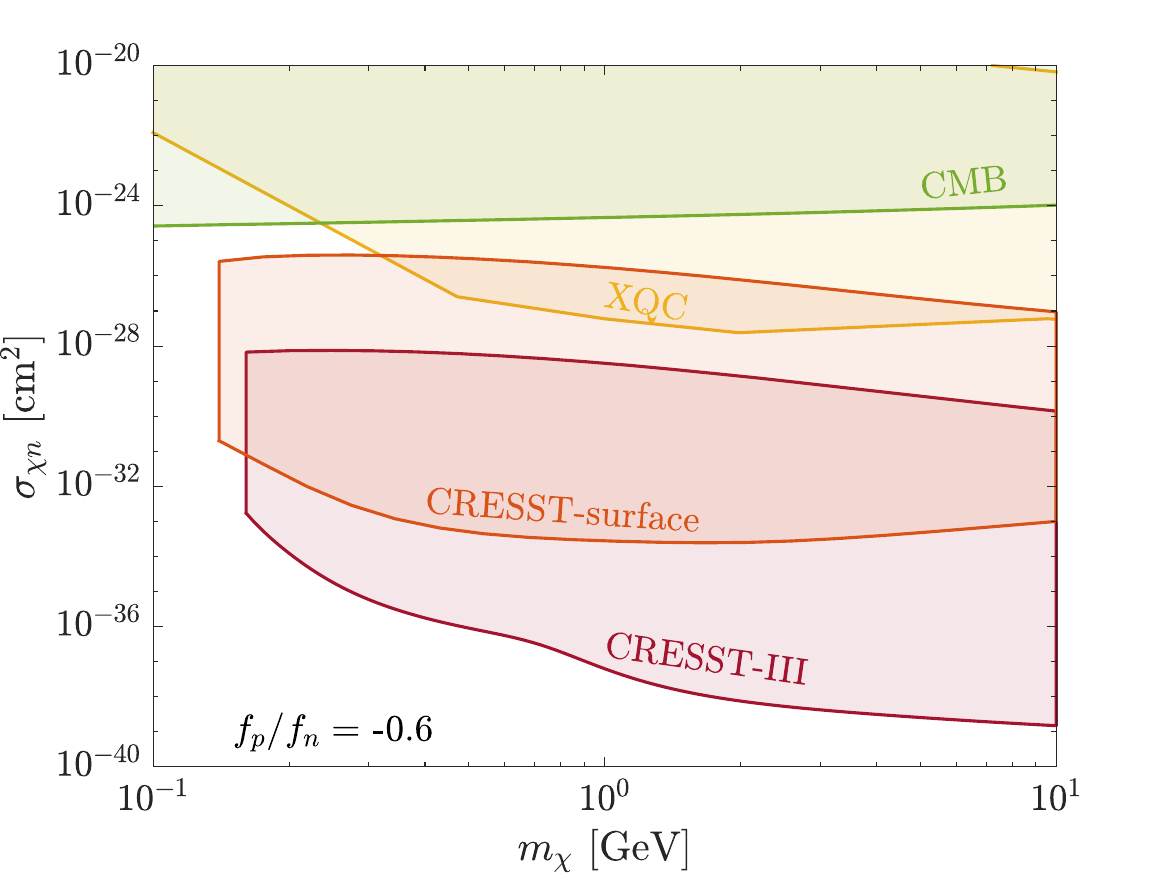}    
    \caption{90\% CL bounds on the dark matter-neutron scattering cross section $\sigma_{\chi n}$ for several values of $f_p/f_n$ with $f_n=1$. The CMB and XQC bounds are obtained by rescaling the corresponding bounds in Fig.~\ref{fig:results_fntofp}.}
    \label{fig:results_fptofn}
\end{figure*}

\textit{\textbf{CRESST surface run.}} The CRESST surface run was operated above ground in 2017 at the Max-Planck-Institute for Physics in Munich, Germany~\cite{CRESST:2017ues}. The prototype cryogenic calorimeter consisted of 0.49~g of Al$_2$O$_3$ with energy resolution $\sigma_d=3.74$~eV. 511 events were observed during the live-time of 2.27~hours. The detector was shielded from above by about 30~cm of concrete roof, 1~mm copper and the atmosphere.

\textit{\textbf{CRESST-III.}} The CRESST-III experiment was operated in the Laboratori Nazionali del Gran Sasso (LNGS) underground laboratory in Italy~\cite{CRESST:2019jnq}. The laboratory provides about 3600~m of water-equivalent shielding. The detector is a 
cryogenic scintillating calorimeter comprised of a 23.6~g CaWO$_4$ crystal with $\sigma_d=4.6$~eV. 441~events were identified with a total exposure of 3.64 kg$\cdot$days after cuts.

Although there are many other deep underground direct detection experiments, we focus on CRESST-III because 
its low threshold allows it to probe light DM. Other experiments provide 
strong limits for $m_\chi \sim 10~\gev$ in the regime of very small scattering cross sections, below the range of our plots.  The overburden cross sections 
for these experiments lie within the CRESST-surface exclusion region, making their impact on our analysis minimal.

\textit{\textbf{XQC.}} The X-ray Quantum Calorimetry (XQC) Project was a rocket-based experiment launched in 1999~\cite{Erickcek:2007jv}. It collected $100~\s$ of data at altitudes between $165$ and $225~\km$. The calorimeters were made mainly of silicon.

Note that atmospheric tests of DM scattering have also been performed using a balloon-based experiment~\cite{Rich:1987st}.  
The bounds reported by that experiment overlap those of XQC for $f_n=f_p$ and $m_\chi \gtrsim 2~\gev$.  However, 
the treatment of the overburden is not described in sufficient detail to allow generalization to other values of 
$f_n / f_p$, without making additional assumptions (as in Ref.~\cite{Hooper:2018bfw}). We therefore do not analyze the data from this experiment.

The 90\% CL constraints on $\sigma_{\chi p}$ are shown in Fig.~\ref{fig:results_fntofp}. To derive the overburden cross sections for the CRESST surface run and CRESST-III,
we modified \texttt{DMprop} to include isospin-violating interactions. 
The lower boundaries are barely affected by the shielding effect.
We also rescale the 90\% CL region excluded by XQC according to the value of $f_n/f_p$.
The floor of the region is scaled up by a factor of $1/R_D$ because of suppressed interactions in the detector, while the ceiling is scaled up by a factor of $1/R_O$ because of the reduced overburden in the atmosphere. The 95\% CL cosmic microwave background (CMB)  limit relies on $\chi$-proton scattering at the time of recombination, so the constraint on $\sigma_{\chi p}$ is independent of $f_n/f_p$. Note that observations of Milky Way satellite galaxies~\cite{nadler2021constraints} and the Lyman-alpha forest~\cite{rogers2022limits}, though subject to cosmological and astrophysical uncertainties, may set stronger constraints than the CMB.

As can be seen from Fig.~\ref{fig:results_fntofp}, isospin-violating interactions shift the constrained regions to higher cross sections. This is understood from Fig.~\ref{fig:Rfactors}, which shows that $R_D \simeq R_O$ for most experimental targets. For $f_n=f_p$, there remains an unconstrained window between the CRESST-surface and XQC/CMB bounds for $m_\chi\lesssim$~0.3~GeV. The window shrinks as  $f_n/f_p$ is lowered until it finally closes for $f_n/f_p\lesssim -0.8$. For $f_n/f_p=-1$, a small window reopens for $m_\chi\gtrsim 6$~GeV, as the XQC detection capability is significantly diminished for its silicon target with nearly equal numbers of protons and neutrons.
Note, however, that the balloon experiment of Ref.~\cite{Rich:1987st} could potentially constrain this open region of parameter space.  

We show 90\% CL constraints on $\sigma_{\chi n}$ in Fig.~\ref{fig:results_fptofn}. Since we consider $|f_p / f_n| \leq 1$, these constraints complement 
those presented in Fig.~\ref{fig:results_fntofp}.
The CMB constraint in Fig.~\ref{fig:results_fptofn} is obtained by recasting the constraint on $\sigma_{\chi p}$ in Fig.~\ref{fig:results_fntofp} by using the relation, $\sigma_{\chi n}^{\rm CMB}=\sigma_{\chi p}^{\rm CMB}/(f_p/f_n)^2$. 
It is noteworthy that for the case of neutron-only scattering ($f_p=0$), CMB constraints are considerably weakened, which opens up an unconstrained window above the XQC excluded region
for the entire $m_\chi$ range of interest and large $\sigma_{\chi n}$.

\section{Conclusions}
\label{sec:Conclusion}

We considered the effect of isospin-violating dark matter interactions on direct detection 
constraints in the regime of light DM and large scattering 
cross section.  In this regime, the effects of isospin-violation are complex, because it 
can cause a reduction in the scattering cross section in the target, but also a reduction in the 
attenuation of the DM flux in the overburden.
For our analysis, we determined the DM velocity distribution at the detector  
using the analytical approach of Ref.~\cite{Cappiello:2023hza}.  This approach, though more complicated 
and time-intensive than the straight-line approximation, correctly incorporates changes to the dark matter 
particle's path through the overburden as a result of scattering, and accounts for statistical fluctuations in the 
energy loss per unit distance.  

We find that if DM-proton and DM-neutron interactions largely cancel, then unconstrained parameter 
space for sub-GeV isopsin-conserving DM gets ruled out because of reduced attenuation in the 
overburden.  On the other hand, windows in parameter mass at higher mass open because the 
sensitivity of XQC is reduced.  If DM couples mainly to neutrons, then constraints from 
CMB measurements are significantly weakend, and parameter space for large $\sigma_{\chi n}$ becomes allowed.

We focused on regions of parameter space in which only a few events are expected at the detector because of attenuation in the overburden. It is then computationally impractical to 
perform a Monte Carlo simulation of such a small flux of 
particles. However, future experiments may explore regions of parameter space in which many 
scattering events are expected, making it possible to study daily modulation with directional detection.  
This large flux regime lends itself to a complete Monte Carlo analysis, since as we have seen,
isospin violation can have a substantial effect on the velocity distribution at the detector.  A 
study of the effect of isospin violation on daily modulation is an interesting topic of future work.

\section*{Acknowledgements}
JK and DM are supported in part by DOE grant DE-SC0010504.
NS is supported by the National Natural Science Foundation of China (NSFC) Project No. 12047503.

\appendix

\bibliography{IVDM}
\end{document}